\documentclass[10pt]{article}
\usepackage[T1]{fontenc}
\usepackage[utf8]{inputenc}
\usepackage{graphicx}
\usepackage{listings}
\usepackage{bm}
\usepackage{latexsym}
\usepackage{amsmath}
\usepackage{amssymb}
\usepackage{booktabs}
\usepackage{array}
\usepackage[UKenglish]{babel}
\usepackage[a4paper,top=3.5cm,bottom=3.5cm,left=3.5cm,right=3.5cm]{geometry}
\usepackage{lineno}
\usepackage{tabularx}
\usepackage{longtable}
\RequirePackage[colorlinks=true
,urlcolor=blue
,anchorcolor=blue
,citecolor=blue
,filecolor=blue
,linkcolor=blue
,menucolor=blue
,pagecolor=blue
,linktocpage=true
,pdfproducer=medialab
]{hyperref}

\usepackage{makeidx}
\newcommand{\be}{\begin{equation}}  
\newcommand{\ee}{\end{equation}}  
\newcommand{\bea}{\begin{eqnarray}}  
\newcommand{\eea}{\end{eqnarray}}  
\makeindex

\makeatletter
\g@addto@macro\bfseries{\boldmath}
\makeatother

\newcommand{\MYhref}[3][blue]{\href{#2}{\color{#1}{#3}}}

\begin{document}

%\linenumbers

\begin{center}
{\LARGE \bf Recent developments and latest results from the xFitter project}

\par\vspace*{2.5mm}\par

{

\bigskip

\large \bf Francesco Giuli\footnote{E-Mail: \MYhref{francesco.giuli@cern.ch}{francesco.giuli@cern.ch}} (on behalf of the xFitter Developers' Team)}

\vspace*{2.5mm}

{CERN, EP Department, CH-1211 Geneva 23, Switzerland}

\vspace*{2.5mm}

{\it Presented at DIS2022: XXIX International Workshop on Deep-Inelastic Scattering and Related Subjects, Santiago de Compostela, Spain, May 2-6 2022}

\vspace*{2.5mm}

\end{center}

\begin{abstract}
In this proceeding, the \texttt{xFitter} project is presented. \texttt{xFitter}  is an open-source package that provides a framework for the determination of the parton distribution and fragmentation functions for many different kinds of analyses in Quantum Chromodynamics. \texttt{xFitter} version 2.2.0 has recently been released and offers an expanded set of tools and options. \texttt{xFitter} has been used for a number of analyses performed recently. An emphasis is given on the recently published study performed by the \texttt{xFitter} developers’ team of the pion fragmentation functions.
\end{abstract}
 
\section{Overview of the project}
The \texttt{xFitter} project~\cite{Alekhin:2014irh} (formerly \texttt{HERAfitter}) is an open-source fitting framework for the determination of parton distribution functions (PDFs) of the proton and light mesons (Ref.~\cite{Novikov:2020snp} illustrates this for the case of pion PDF). Version 2.2.0\footnote{Available on the web at:~\url{https://www.xfitter.org/xFitter/xFitter/DownloadPage/}} has been recently released and offers a wide and comprehensive set of tools and options for the determination of parton densities. Various data sets from fixed-target experiments, as well as from $ep$ (HERA), $p\bar{p}$ (Tevatron) and $pp$ (LHC) colliders, are already incorporated in \texttt{xFitter}, and they can be fitted using predictions up to next-to-next-to-leading order (NNLO) in quantum chromodynamics (QCD) and next-to-leading order (NLO) in electroweak (EW) perturbation theory. Several theoretical calculations are available in this framework, as well as plotting tools to allow users to better visualize the results. The framework can be used to study the impact of new measurements from hadron colliders on PDFs, check the consistency of measured data from different experiments, test different theoretical assumptions, and also assess the impact of future colliders (see Ref.~\cite{LHeC:2020van} for LHeC and Ref.~\cite{Accardi:2012qut} for EIC pseudo-data respectively).\\
More than 100 publications have already used the \texttt{xFitter} framework, thanks to its modular structure which allows for various theoretical and methodological options. Interfaces to \texttt{HATHOR}~\cite{Aliev:2010zk}, \texttt{FastNLO}~\cite{Britzger:2015ksm}, \texttt{APFEL}~\cite{Bertone:2013vaa,Bertone:2017gds}, \texttt{APFELgrid}~\cite{Bertone:2016lga}, \texttt{APPLgrid}~\cite{Carli:2010rw}, \texttt{LHAPDF}~\cite{Buckley:2014ana} and \texttt{QCDNUM}~\cite{Botje:2010ay} are already contained in the platform. More detailed documentation can be found in Ref.~\cite{Alekhin:2014irh}, and an overview of available tutorial can be found in Ref.~\cite{xFitterDevelopersTeam:2017xal}.

\section{Recent and future developments}
A significant restructuring of the code is present in the latest \texttt{xFitter} release - the new version providing significantly improved modularity for PDF parametrization, evolution, theory predictions and minimization. This additional flexibility simplifies further developments of the code. The existing functionalities have been improved and the code capabilities expanded. The new release is also characterized by the usage of modern, industry-standard libraries, with several improvements of the execution time required for minimisation by more than one order of magnitude. Moreover, \texttt{xFitter} can be interfaced to modern minimization packages, such as \texttt{CERES}, providing additional opportunities to employ more sophisticated and faster automatic differentiation.\\
Further developmenst to allow functionality beyond that of the determination of conventional PDFs have already been planned by the xFitter Developers' Team. For example, the functionality for fitting PDFs and Fragmentation Functions (FFs) will be introduced. Furthermore, transverse-momentum-dependent (TMD) parton
distribution functions~\cite{Angeles-Martinez:2015sea} are being developed by including branching scale-dependent resolution parameters~\cite{Hautmann:2019biw} as a new functionality. TMDs will
be introduced independently of the parton branching method using the existing interface to the \texttt{DYTURBO}~\cite{Camarda:2019zyx} package. Moreover, since one of the main targets of state-of-the-art PDF global fits is to include theoretical uncertainties in the PDF (see the discussion in Ref.~\cite{ATLAS:2021vod}), \texttt{xFitter} plans to provide PDF sets incorporating theoretical uncertainties by applying the
resummation-scale technique presented in Ref.~\cite{Bertone:2022sso}.

\section{The pion fragmentation function}
\begin{figure}[t!]
\begin{center}
\includegraphics[width=0.9999\textwidth]{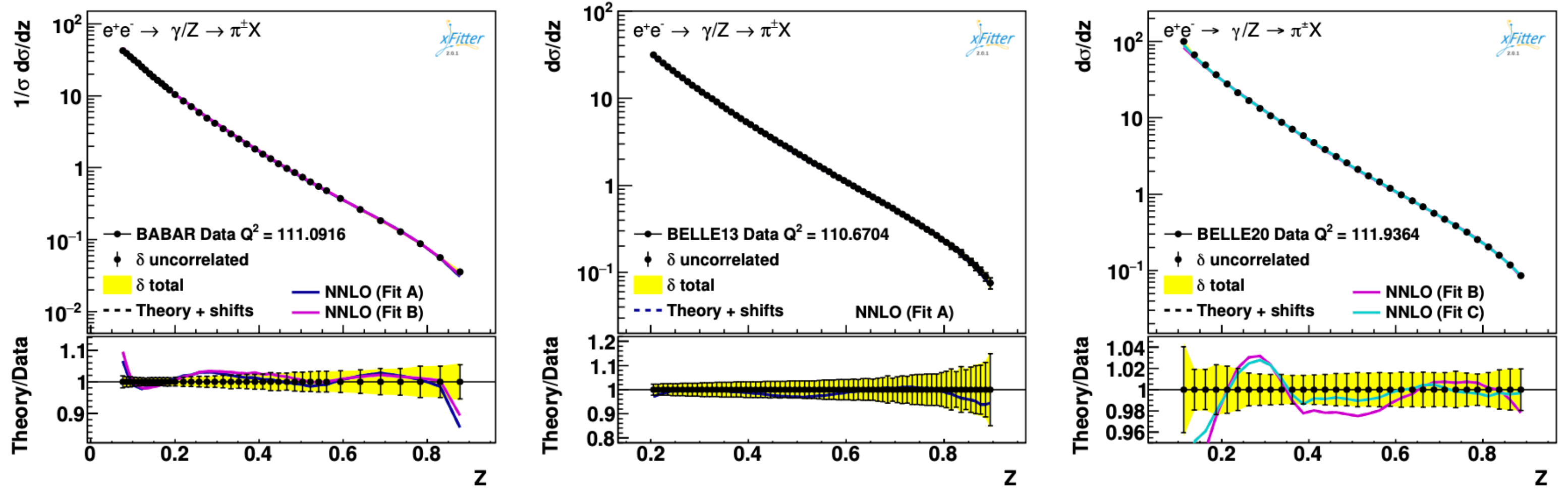}
\end{center}
\caption{A comparison of BELLE13, BELLE20, and BaBar for fits A, B, and C. Note that not all data sets are included in each fit. These plots are taken from Ref.~\cite{Abdolmaleki:2021yjf}.} 
\label{fig:data}
\end{figure}
The first open-source analysis of FFs of charged pions performed at both NLO and NNLO accuracy in perturbative QCD is presented was presented by the \texttt{xFitter} Collaboration in Ref.~\cite{Abdolmaleki:2021yjf} and is summarised here. The resulting fit is called ``IPM-XFITTER''. This study incorporates an up-to-date and comprehensive set of pion production data from single-inclusive annihilation ($e^{+}e^{-}\rightarrow\pi^{\pm}X$, SIA) processes, together with the most recent measurements of inclusive cross-sections of single pion by the BELLE collaboration. These data contain inclusive and flavour-tagged measurements of pion scaled energy,$z=2E_{\pi^{\pm}}/\sqrt{s}$, calculated at different center of mass energies of $\sqrt{s}=Q$ with charged pion energy $E_{\pi^{\pm}}$, corresponding to individual charm, or bottom quarks, and the sum of light quarks.\\
The $z$-dependence of the FFs at an initial scale $Q_{0}^{2} =$ 25~GeV$^{2}$ is parametrised as follows:
\begin{equation}
D_{i}^{\pi^{\pm}}(z,Q_{0})=\dfrac{\mathcal{N}_{i}z^{\alpha_{i}}(1-z)^{\beta_{i}}[1+\gamma_{i}(1-z)^{\delta_{i}}]}{B[2+\alpha_{i},\beta_{i}+1]+\gamma_{i}B[2+\alpha_{i},\beta_{i}+\delta_{i}+1]}
\end{equation}
where $B[a,b]$ is the Euler beta function and each parton flavour has five free parameters. The fitted flavour combination are $i=u^+,d^+,s^+,c^+,b^+$ and $g$.\\
To investigate the influence if the BELLE13~\cite{Belle:2013lfg}, BELLE20~\cite{Belle:2020pvy} and BaBar~\cite{BaBar:2013yrg} data sets on FFs was a primary goal of this QCD analysis. Thus, five different sets of fits have been performed and they are summarised below:
\begin{itemize}
\item \textbf{Fit A}: This fit focuses on the impact of the BELLE13 data set. Thus, the BELLE20 data are excluded.
\item \textbf{Fit B}: This fit focuses on the impact of the BELLE20 data set. Thus, the BELLE13 data are excluded.
\item \textbf{Fit C}: This fit focuses on the impact of the BELLE20 data without the BaBar set. Thus, the BaBar and BELLE13 data are excluded.
\item \textbf{Fit D}: This fit focuses on the impact of cutting the low-$z$ BELLE20 data. Thus, the BaBar and BELLE13 data are excluded and a cut, $z>$ 0.2, is imposed on the BELLE20 data.
\item \textbf{Fit E}: This fit focuses on the impact of the BELLE20 and BaBar sets with cuts imposed to remove low-$z$ data. Thus,the BELLE13 data are excluded and a cut, $z>$ 0.2, is imposed on the BELLE20 data, and $z >$ 0.1 on the BaBar data.
\end{itemize}
\begin{figure}[t!]
\begin{center}
\includegraphics[width=0.9999\textwidth]{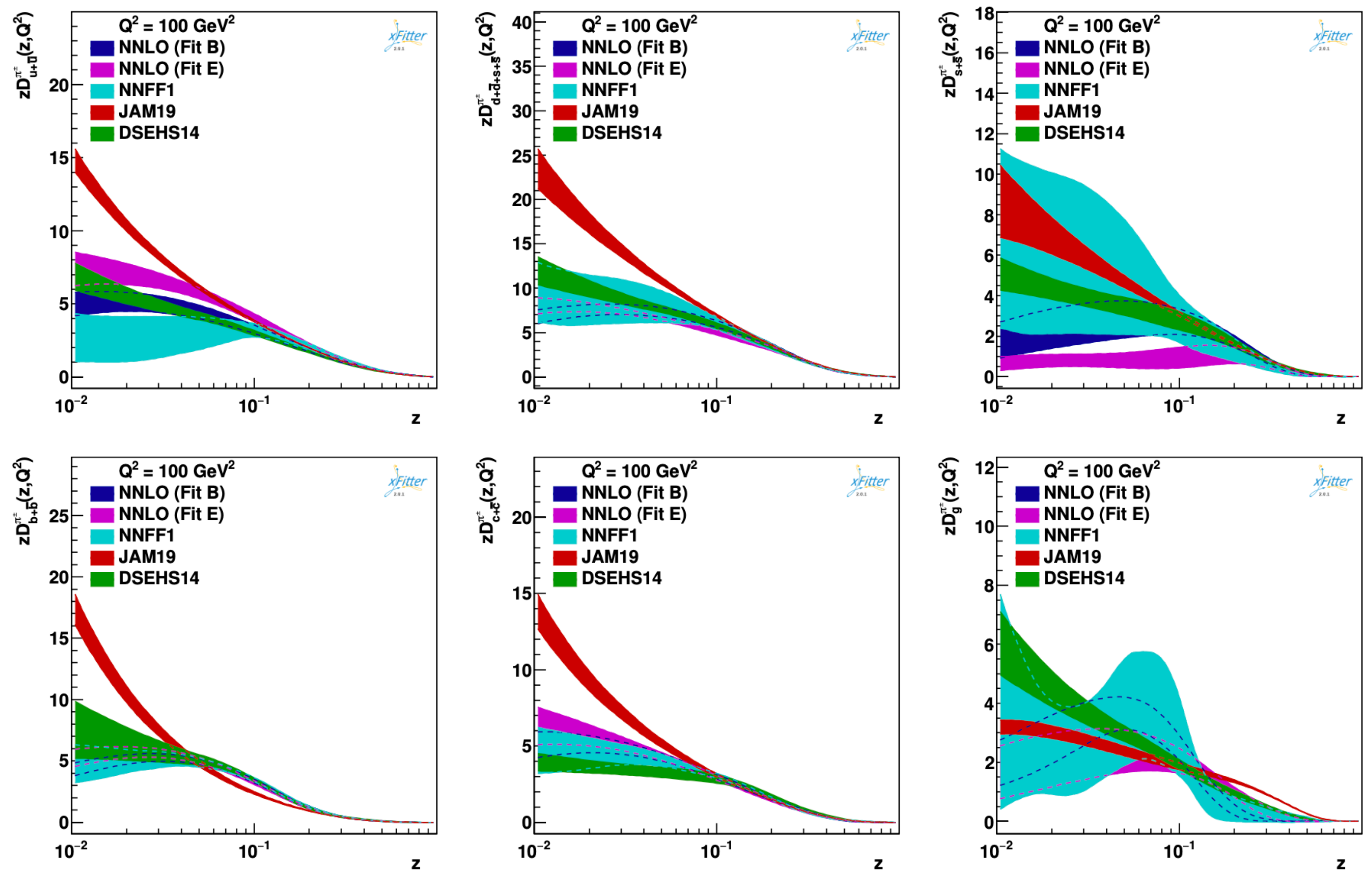}
\end{center}
\caption{A comparison of fits E and B for charged pion FFs ($\pi^{+}+\pi^{-}$) at NNLO with results from the literature at $Q^{2}$ = 100 GeV$^{2}$. The predictions from DSEHS14~\cite{deFlorian:2014xna} and JAM19~\cite{Sato:2019yez} at NLO, NNFF1~\cite{Bertone:2017tyb} at NNLO, with their uncertainties, are also displayed. These plots are taken from Ref.~\cite{Abdolmaleki:2021yjf}.} 
\label{fig:FFs}
\end{figure}
The comparison between our fits A, B and C with the data for BELLE13, BELLE20 and BaBar is displayed in Figure~\ref{fig:data}. As regards the BELLE13 data, within the large uncertainties Fit A is consistent with the data; fits B and C provide a good description of the BaBar or BELLE data with the exception of the low-$z$ region, where the two data sets pull the fits in opposite directions. Because of this poor description at low $z$, additional selection in $z$ was considered for fits D and E.\\
Comparisons of fits E and B are displayed in Figure~\ref{fig:FFs} along with results from the literature. The preferred fit is Fit E, with Fit B chosen to highlight the impact of low-$z$ cuts. Comparison is shown with results from the DSEHS14~\cite{deFlorian:2014xna}, JAM19~\cite{Sato:2019yez} and NNFF1~\cite{Bertone:2017tyb,Nocera:2017qgb,Bertone:2017xsf} collaborations which are computed either at NLO (DSEHS14 and JAM19) or at NNLO (NNFF1). Figure~\ref{fig:FFs} shows these results at $Q^{2}$ = 100 GeV$^{2}$. The up and down FFs are generally compatible with NNFF1 and DSEHS14 at larger $z$, but they differ in the low-$z$ region, and similar conclusions can be applied to the $c^{+}$ and $b^{+}$ heavy quarks.  
For the strange quark, results have a larger spread, suggesting an overall increased uncertainty, in part reflecting the minimal constraint imposed on the strange FF by the chosen data. Comparing the \texttt{xFitter} gluon FF with NNFF1 and DSEHS14, again it is seen that the FFs are generally compatible, lying within the NNFF1 uncertainty and overlapping with the DSEHS14 uncertainty for the larger $z$ region.\\
It is interesting to notice that the JAM19 FFs show a different behaviour in contrast to the above-mentioned analyses. The FFs from JAM19 have a much steeper slope at small $z$ for the quark flavours, while the gluon generally is generally higher than the \texttt{xFitter} results for intermediate to larger $z$ values. The JAM19 uncertainty bands are much smaller in comparison with the \texttt{xFitter} result and the NNFF1 analysis. This can be understood as the JAM19 focus which was on SIDIS in the region $z\gtrsim$ 0.2, and in this region there is closer agreement among the FFs. Moreover, the JAM19 collaboration obtain their results by a simultaneous determination of both PDFs and FFs, so it would be interesting to investigate further to determine to what extent any difference might be separately attributable to this combined analysis and the choice of data sets.

\section{Conclusion}
The \texttt{xFitter} 2.2.0 program is a versatile, flexible, modular, and
comprehensive tool that can facilitate analyses of experimental data
with a variety of theoretical calculations. This analysis exemplifies the capability of \texttt{xFitter} to analyse both PDFs and FFs of both hadrons (e.g., protons and nuclei) and mesons (e.g. pions). This is facilitated by incorporating experimental data from a wide range of experiments, and implementing NNLO theoretical calculations in perturbation theory. We encourage the use of \texttt{xFitter}, and welcome new contributions from the community as we continue to incorporate the latest theoretical advances and precision experimental data.

\end{document}